\documentclass[aps,prresearch,twocolumn,floatfix,superscriptaddress,longbibliography,nofootinbib]{revtex4-2}

\usepackage{graphicx}
\usepackage{amsmath,amssymb,bm}
\usepackage{float}
\makeatletter 
\let\float@end\float@end@ltx\let\newfloat\newfloat@ltx
\makeatother
\usepackage{algorithm}
\usepackage{algorithmic}
\usepackage{xcolor}
\usepackage{xr-hyper}
\usepackage[colorlinks=true,linkcolor=blue!60!black,citecolor=blue!60!black,urlcolor=blue!60!black]{hyperref}

\newcommand{\Var}{\mathrm{Var}}
\newcommand{\Lmin}{L^2_{\min}}

\begin{document}

\title{Pool-free, residual-triggered ansatz growth for variational imaginary-time evolution:\\ when a fixed operator order beats adaptive screening}

\author{Yu-Hao Chen}
\email{d13222001@ntu.edu.tw}
\affiliation{Department of Physics, National Taiwan University, Taipei, Taiwan}
\affiliation{Department of Artificial Intelligence, Chang Gung University, Taoyuan, Taiwan}

\author{Renata Wong}
\email{renata.wong@cgu.edu.tw}
\thanks{Corresponding author.}
\affiliation{Department of Artificial Intelligence, Chang Gung University, Taoyuan, Taiwan}
\affiliation{Department of Neurology, Chang Gung Memorial Hospital, Keelung, Taiwan}
\affiliation{Artificial Intelligence Research Center, Chang Gung University, Taoyuan, Taiwan}

\date{\today}

\begin{abstract}
Adaptive variational algorithms such as ADAPT-VQE and AVQITE screen an operator pool at every growth step, and ADAPT-VQE also re-optimizes all parameters. We study a pool-free alternative for variational imaginary-time evolution (VITE). Its layers are fixed in advance by nested commutators of the Hamiltonian with a parent operator of the starting state, grouped by symmetry and particle number. The next layer is appended when the relative McLachlan residual exceeds a threshold. We ask whether this dynamic VITE reaches the accuracy of adaptive methods at a comparable or lower measurement cost. To make the comparison fair, every method starts from the classically computed state that suits it best. At a relative error of $10^{-4}$, or at chemical accuracy for molecules, dynamic VITE needs 1.8--6.0 times fewer circuits than the cheapest adaptive variant on seven of thirteen tested systems. 
It is more expensive on transverse-field Ising models, where adaptive methods need few operators, and at tight targets, where the metric of its deeper circuits dominates its cost. On the Hubbard chains the choice of start matters most: from the restricted Hartree-Fock state, with the hopping term as parent, dynamic VITE needs 25 times fewer circuits on the four-site chain than from the N\'eel state, and on the six-site chain it reaches $10^{-6}$ instead of stalling at $2\times10^{-3}$.
\end{abstract}

\maketitle

\section{Introduction}

Imaginary-time evolution $|\psi(\tau)\rangle\propto e^{-H\tau}|\psi_0\rangle$ converges to the ground state of $H$ whenever the initial state overlaps with it. Variational imaginary-time evolution (VITE)~\cite{McArdle2019,Yuan2019} replaces the non-unitary flow by the trajectory of a parametrized circuit, with the parameter velocities fixed by McLachlan's variational principle~\cite{McLachlan1964}. Each step requires the metric $M$ and force $V$ of the circuit, estimated with Hadamard-test or direct-measurement circuits~\cite{McArdle2019,Mitarai2019}. If the tangent space of the circuit cannot represent the exact direction, the trajectory stalls above the ground state.

McLachlan's distance at its minimum, $\Lmin$, measures this deficiency~\cite{Yuan2019}. Adaptive variational quantum imaginary-time evolution (AVQITE)~\cite{Gomes2021} and its real-time counterpart~\cite{Yao2021} grow the circuit when $\Lmin$ exceeds a cutoff, screening an operator pool for the operator that lowers $\Lmin$ most. ADAPT-VQE~\cite{Grimsley2019,Tang2021} screens a pool for the largest energy gradient and re-optimizes all parameters after each growth step. Both pay a screening cost at every growth event, and ADAPT-VQE also pays for the optimizer. On hardware these costs are measured in circuits. Quantum imaginary-time evolution (QITE)~\cite{Motta2020} instead fits a unitary to each imaginary-time step. Recent variants lower its cost by regularizing the generators toward sparsity~\cite{GomezLurbe2026} or by adaptive time steps and circuit compression~\cite{AcevedoMelendez2026}.

We separate the two decisions these methods make at each growth event: \emph{when} to grow and \emph{what} to add. The \emph{what} is fixed in advance by a schedule of layers built from nested commutators of $H$ with a parent operator of the starting state, split into classes by the symmetry group of the problem and, for fermions, by particle-number conservation. Nested commutators have been used as counterdiabatic terms~\cite{Hegade2022,Chandarana2022} and as an operator pool for ADAPT-VQE~\cite{Tancara2026}; here they replace operator selection altogether. The \emph{when} is decided by the relative residual $\rho=\Lmin/\Var(H)\in[0,1]$, the fraction of the exact imaginary-time direction that the circuit misses. The next layer is appended when $\rho$ exceeds a threshold $\epsilon$. We call the result dynamic VITE.

Our question is whether this pool-free strategy reaches the accuracy of the adaptive methods at a comparable or lower measurement cost. A fair answer needs a fair start. The best starting state depends on the method: a mean-field state that is cheap to improve for one method may stall another. We therefore run every method from each of a few classically computed candidate states, such as the optimized product state of a spin model and the Hartree--Fock determinants of a Hubbard chain, and report each method from the start with which it is cheapest. We compare dynamic VITE with AVQITE and ADAPT-VQE using the same operators, and with stronger variants using finer pools and batched growth~\cite{Tang2021,Yordanov2021,Anastasiou2024}. Every method is charged the number of distinct measurement circuits. The thirteen systems are transverse-field and mixed-field Ising chains and tori, Fermi--Hubbard chains~\cite{Lieb1968}, and the molecules H$_4$, LiH and H$_6$, with 6--16 qubits.

The main results are as follows. At a relative error of $10^{-4}$ or at chemical accuracy, dynamic VITE is cheaper than every adaptive variant by factors of 1.8--6.0 on seven of the thirteen systems, and within about 20\% of the cheapest one on three more. Its saving comes from avoiding screening and re-optimization. It requires a short schedule, which a large symmetry group, a good parent operator, or, for molecules, an order taken from classical coupled-cluster amplitudes provides. Dynamic VITE is more expensive where adaptive methods need few operators and at tight targets. Further results, including all starting states, seeded ensembles, size scaling, a longitudinal-field scan, ablations and hybrid variants, are given in the Supplemental Material~\cite{SM}.

\section{Method}\label{sec:method}

\subsection{McLachlan residual}\label{sec:vite}

The normalized imaginary-time equation is $\partial_\tau|\psi\rangle=-(H-E_\tau)|\psi\rangle$ with $E_\tau=\langle\psi|H|\psi\rangle$. For a circuit $|\phi(\bm\theta)\rangle=U_L(\theta_L)\cdots U_1(\theta_1)|\psi_0\rangle$ with real parameters and projected tangent vectors $|t_i\rangle=(1-|\phi\rangle\langle\phi|)\,\partial_{\theta_i}|\phi\rangle$, McLachlan's principle minimizes
\begin{equation}
L^2(\dot{\bm\theta})=\Big\|\sum_i\dot\theta_i|t_i\rangle+(H-E_\tau)|\phi\rangle\Big\|^2 ,
\label{eq:L2}
\end{equation}
which gives $M\dot{\bm\theta}=V$ with $M_{ij}=\mathrm{Re}\langle t_i|t_j\rangle$ and $V_i=-\mathrm{Re}\langle t_i|(H-E_\tau)|\phi\rangle$. With $\dot{\bm\theta}=M^{+}V$ and the energy variance $\Var(H)$,
\begin{equation}
\Lmin=\Var(H)-V^{\mathsf T}M^{+}V,\qquad \rho=\frac{\Lmin}{\Var(H)}\in[0,1] .
\label{eq:Lmin}
\end{equation}
Appending a layer at zero angle leaves the state and $\Var(H)$ unchanged and enlarges the tangent space, so $\rho$ cannot increase at the moment of growth. At a stationary point of the variational flow $\rho=1$, and along an exactly followed trajectory $\rho=0$ (see Ref.~\cite{SM}).

\subsection{Residual-triggered growth}\label{sec:growth}

Algorithm~\ref{alg:dvite} takes an ordered schedule of $K$ layers $G_k$. The circuit starts empty, so $\rho=1$ at $\tau=0$ and the first layer is appended at once. No pool is screened, and the ground-state energy is never used. We use $\epsilon=0.9$: a layer is added when the circuit captures less than 10\% of the exact descent rate.

\begin{algorithm}[t]
\caption{Dynamic VITE: pool-free, residual-triggered growth.}\label{alg:dvite}
\begin{algorithmic}[1]
\REQUIRE schedule $(G_1,\dots,G_K)$; start $|\psi_0\rangle$; threshold $\epsilon$; step $\delta\tau$; step limit $s_{\max}$; tolerance $\eta$
\STATE $\bm\theta\leftarrow()$, $k\leftarrow1$
\FOR{$s=0,1,2,\dots$}
  \STATE estimate $E$, $\langle H^2\rangle$, $V$, $M$; compute $\rho$ [Eq.~\eqref{eq:Lmin}]
  \WHILE{$\rho>\epsilon$ \AND $k\le K$}
     \STATE append $e^{-i\theta_k G_k}$ with $\theta_k=0$; $k\leftarrow k+1$; update $M$, $V$, $\rho$
  \ENDWHILE
  \STATE $h\leftarrow\delta\tau$; \textbf{while} $E(\bm\theta+hM^+V)>E(\bm\theta)$: $h\leftarrow h/2$
  \STATE $\bm\theta\leftarrow\bm\theta+h\,M^{+}V$
  \STATE \textbf{stop} if $\Var(H)<\eta$ or $s=s_{\max}$
\ENDFOR
\end{algorithmic}
\end{algorithm}

\begin{table*}[t]
\caption{Benchmark systems and starting states. $N_q$: qubits; $|\mathcal G|$: order of the symmetry group; $K$: scheduled layers; strings: Pauli strings per layer; overlap: $|\langle\psi_0|\psi_{\rm gs}\rangle|$ for each candidate start, in the order listed. Where $|\mathcal G|$ or the string counts depend on the start, they are given in the same order. For the TFIM ring the tilted state coincides with $|{-}\rangle^{\otimes N}$.}\label{tab:systems}
\begin{ruledtabular}
\begin{tabular}{llcccccc}
Class & System & $N_q$ & starts & $|\mathcal G|$ & $K$ & strings & overlap\\
\hline
1D & TFIM ring, $J=h_x=1$, $N=6$--16 & 6--16 & $|{-}\rangle$ & $2N$ & $N/2+12$ & $N$, $2N$ & 0.52 ($N=12$)\\
1D & Mixed-field Ising ring, $N=8$--16 & 8--16 & $|{-}\rangle$, tilted & $2N$ & 40 & $N$, $2N$ & 0.16, 0.99 ($N=12$)\\
2D & TFIM, $3\times3$ torus & 9 & $|{-}\rangle$, tilted & 72 & 40 & 36--72; 6--72 & 0.85, 0.72\\
2D & TFIM, $3\times4$ torus & 12 & $|{-}\rangle$, tilted & 48 & 40 & 24--48; 4--48 & 0.82, 0.67\\
2D & Mixed-field Ising, $3\times3$ torus & 9 & $|{-}\rangle$, tilted & 72 & 40 & 6--72 & 0.55, 0.98\\
2D & Mixed-field Ising, $3\times4$ torus & 12 & $|{-}\rangle$, tilted & 48 & 40 & 4--48 & 0.44, 0.98\\
Fermionic & Hubbard chain, $L=4$, $U/t=4$ & 8 & N\'eel, RHF, UHF & 2, 4, 2 & 40 & 4--16; 8--32; 4--16 & 0.46, 0.85, 0.66\\
Fermionic & Hubbard chain, $L=6$, $U/t=4$ & 12 & N\'eel, RHF, UHF & 2, 4, 2 & 80 & 4--16; 8--16; 4--16 & 0.33, 0.77, 0.57\\
Fermionic & H$_4$, 1.5 and 2.0 \AA, STO-3G & 8 & RHF & 1 & 40 & 2--8 & 0.87, 0.69\\
Fermionic & LiH, 1.6 and 3.0 \AA, STO-3G & 12 & RHF & 1 & 80 & 2--8 & 0.99, 0.85\\
Fermionic & H$_6$, 1.5 \AA, STO-3G & 12 & RHF & 1 & 160 & 2--8 & 0.80\\
\end{tabular}
\end{ruledtabular}
\end{table*}

\subsection{Commutator schedule}\label{sec:schedule}

Let $H$ be real, and let the start $|\psi_0\rangle$ be the ground state of a real parent operator $H_0$. Define
\begin{equation}
C_1=i[H,H_0],\qquad C_{n+1}=-[H,[H,C_n]] .
\label{eq:Cn}
\end{equation}
Every $C_n$ is a real combination of Pauli strings with an odd number of $Y$, whose exponentials keep real states real (see~\cite{SM}). Each level $C_n$ is split into classes of Pauli strings related by the symmetry group $\mathcal G$, the qubit permutations that leave $H$ and $H_0$ invariant, found automatically from candidate lattice operations. For number-conserving fermionic problems, strings that differ only by $X\leftrightarrow Y$ on the same qubits are also merged, so that every layer conserves the spin-up and spin-down particle numbers. Each class not seen at an earlier level becomes one layer,
\begin{equation}
G_c=\sum_{P\in c}\frac{a_P}{\max_{P'\in c}|a_{P'}|}\,P,
\label{eq:layer}
\end{equation}
with $a_P$ the coefficient of $P$ in $C_n$. Layers are ordered by level, then, on lattices, by support size and coefficient magnitude, and for molecules by coefficient magnitude. We keep the first $K$ layers.

The order has a physical basis. Acting on the start, $C_1$ moves it along the exact imaginary-time direction $-(H-E_{\rm ref})|\psi_0\rangle$, with each component weighted by its excitation energy under $H_0$. When each level-1 class reaches a single eigenspace of $H_0$, the classes span the exact direction and $\rho=0$ at $\tau=0$. The higher levels are the terms of the nested-commutator expansion of the adiabatic gauge potential that transports the ground state of $H_0$ to that of $H$~\cite{Kolodrubetz2017,Sels2017,Claeys2019}, with one angle per class in place of its coefficients (see~\cite{SM}).

\emph{Classically ordered molecular schedules.} For a molecule with the Hartree--Fock start and the Fock operator as parent, each level-1 class is a set of number-conserving excitations weighted by orbital-energy differences, which favors excitations out of core orbitals. We therefore also rank the first-level classes by the summed absolute amplitudes of a classical coupled-cluster singles and doubles (CCSD) calculation~\cite{Sun2020} for the determinants they create from the Hartree--Fock state ($|t_i^a|$ and $|t_{ij}^{ab}+t_i^at_j^b-t_i^bt_j^a|$). Unranked classes keep their default order behind the ranked ones. We call this the CCSD order and use it for all molecules. Ref.~\cite{SM} compares it with the default order and with orders from second-order energies and orbital entropies.

\subsection{Starting states and parent operators}\label{sec:starts}

Each method begins in a chosen initial state, and the operator schedule is built using a parent Hamiltonian $H_0$ whose ground state is that initial state. We consider the following candidates, each computed classically and prepared by a fixed circuit with no measurement circuits charged.

\emph{Ising models.} (i)~$|{-}\rangle^{\otimes N}$, the ground state of $H_0=|h_x|\sum_jX_j$, which keeps the spin-flip symmetry of the transverse-field models. (ii)~The tilted product state, the uniform product state of lowest energy, with every spin along a Bloch vector $\bm b=(\sin\vartheta,0,\cos\vartheta)$ found by a one-parameter minimization. Its parent is $H_0=-|h_x|\sum_j(b_xX_j+b_zZ_j)$. At $h_z=0$ it breaks the spin-flip symmetry.

\emph{Hubbard chains.} (i)~The N\'eel state, with a staggered field as parent. (ii)~The restricted Hartree--Fock (RHF) determinant. At half filling its orbitals are the tight-binding orbitals, so its parent is the hopping term of $H$. (iii)~The unrestricted (antiferromagnetic) Hartree--Fock (UHF) determinant, with the UHF mean-field Hamiltonian as parent. The determinants are prepared by Givens-rotation circuits of depth $O(L)$~\cite{Kivlichan2018}.

\emph{Molecules.} The RHF determinant, with the Fock operator as parent. For all five molecular systems the UHF solution coincides with it.

The schedule pools of AVQITE and ADAPT-VQE are rebuilt from the parent of each start. ADAPT-VQE started from the tilted state may also use the schedule of $|{-}\rangle^{\otimes N}$ as pool, which is cheaper on the mixed-field ring. For each method, system and target we then report the cheapest start (and pool variant), chosen in hindsight. This choice favors every method equally. The candidates and the overlaps of their states with the ground state are listed in Table~\ref{tab:systems}.

\section{Systems, baselines and cost}\label{sec:systems}

\emph{Systems} (Table~\ref{tab:systems}). The transverse-field Ising (TFIM) ring $H=J\sum_jZ_jZ_{j+1}+h_x\sum_jX_j$ at $J=h_x=1$, and the mixed-field ring with an added $h_z\sum_jZ_j$ at $(J,h_x,h_z)=(-1,1.05,0.5)$~\cite{Banuls2011}, with the dihedral group of the ring. The TFIM and mixed-field model ($h_z=0.5$) on $3\times3$ and $3\times4$ tori with $J=-1$ and $h_x=3.04$, close to the critical field of the square lattice~\cite{Blote2002}, with the translations and reflections of the torus. The open Fermi--Hubbard chain with $L=4$ and 6 sites, $U/t=4$ and half filling, mapped by the Jordan--Wigner transformation~\cite{JordanWigner1928,Chen2026} with all spin-up modes first. H$_4$ (1.5 and 2.0~\AA), LiH (1.6 and 3.0~\AA) and linear H$_6$ (1.5~\AA) in the STO-3G basis~\cite{Hehre1969}, with integrals from PySCF~\cite{Sun2020} and qubit Hamiltonians from OpenFermion~\cite{McClean2020}. Fermionic energies are compared with the lowest eigenvalue in the sector of the start.

\emph{Baselines.} AVQITE screens every scheduled layer when $\rho>0.9$ and appends the one that lowers $\Lmin$ most, with reuse and a depth cap of $K$; it differs from dynamic VITE only in how the next operator is chosen. ADAPT-VQE appends the pool operator with the largest energy gradient and re-optimizes all parameters with exact gradients and the BFGS method~\cite{Nocedal2006}, until the largest gradient falls below $10^{-5}$ or $K$ operators are reached. Both use the schedule as pool; for molecules ADAPT-VQE also uses the spin-conserving UCCSD pool~\cite{Romero2019}. The stronger baselines are ADAPT-VQE, TETRIS-ADAPT-VQE and AVQITE with finer pools: 2-local single Pauli strings (qubit pool)~\cite{Tang2021} for the spin models, qubit excitations (QEB)~\cite{Yordanov2021} and UCCSD excitations for the molecules, and generalized spin-conserving singles and doubles (GSD)~\cite{Lee2019} and their qubit-excitation form for the Hubbard chains. ADAPT-type costs are sensitive to rounding-level differences in the optimizer, so where they decide a comparison we report medians over four to seven seeded runs. Details are in Ref.~\cite{SM}.

\emph{Cost model.} We count distinct circuits, each run with the same number of shots, using Table~\ref{tab:costmodel} for every method. $g(A)$ is the number of qubit-wise-commuting groups of the Pauli terms of $A$~\cite{Verteletskyi2020}, and $|G_l|$ the number of strings in layer $l$. We compare methods by the circuits needed to first reach a target relative error $(E-E_0)/|E_0|$ of $10^{-4}$ or $10^{-6}$, and for molecules also chemical accuracy, $E-E_0<1.6$~mHa. The exact energy $E_0$ is used only for this evaluation. The count ignores circuit depth and shot allocation, and it does not use symmetry to reduce the circuits of symmetric layers, which disfavors dynamic VITE.

\begin{table}[t]
\caption{Circuits counted per estimated quantity, for all methods.}\label{tab:costmodel}
\begin{ruledtabular}
\begin{tabular}{ll}
Quantity & Circuits \\
\hline
$E=\langle H\rangle$, step-control trial & $g(H)$\\
$\langle H^2\rangle$ (for $\rho$) & $g(H^2)$\\
$V_l$ or $\partial_l E$ & $|G_l|\,g(H)$\\
$M_{ij}$, $i\le j$ & $|G_i|\,|G_j|$\\
AVQITE screening, candidate $c$ & $|G_c|[g(H)+\textstyle\sum_l|G_l|+|G_c|]$\\
ADAPT screening, candidate $c$ & $|G_c|\,g(H)$\\
\end{tabular}
\end{ruledtabular}
\end{table}

\section{Results}\label{sec:results}

\subsection{Every method from its best start}\label{sec:comparison}

\begin{table*}[t]
\caption{Every method from its best start. Circuits needed by dynamic VITE ($\epsilon=0.9$) to first reach each target relative error, and circuits needed by the other methods \emph{relative to} it (values above 1 mean dynamic VITE is cheaper). Each entry uses the start with which that method is cheapest; superscripts mark the tilted product state (t), the RHF (r) and the UHF (u) determinant, and no superscript the original start ($|{-}\rangle^{\otimes N}$, N\'eel or Hartree--Fock). AVQITE and ADAPT use the schedule as pool. UCCSD: ADAPT-VQE with the UCCSD pool (molecules only). Best stronger: the cheapest of the stronger baselines (Sec.~\ref{sec:systems}); on H$_4$ at 1.5~\AA\ at $10^{-4}$ it is AVQITE with the absolute cutoff of Ref.~\cite{Gomes2021} (Sec.~S8 of Ref.~\cite{SM}). Cheapest: the smallest of these ratios, i.e.\ the cheapest adaptive method from its best start. $\epsilon{=}0.1$: dynamic VITE with $\epsilon=0.1$ from the same start. ADAPT-type entries are medians over seeded runs where these exist. ``--'': not reached; empty: not applicable. For molecules, dynamic VITE uses the CCSD order, the targets are chemical accuracy and $10^{-4}$, and the adaptive methods may use the same classical information: superscript c marks the CCSD-ordered schedule as pool, i a UCCSD pool whose operators start from their CCSD amplitudes. Above the double line: systems where dynamic VITE matches or beats the adaptive methods; below it: the transverse-field models.}\label{tab:summary}
\begin{ruledtabular}
\begin{tabular}{llcccccccc}
 & & \multicolumn{2}{c}{dynamic VITE} & \multicolumn{5}{c}{adaptive methods: circuits / dynamic VITE} & dyn.\ VITE \\
System & target & start & circuits & AVQITE & ADAPT & UCCSD & best stronger & \textbf{cheapest} & $\epsilon{=}0.1$ \\
\hline
Mixed-field Ising 1D, $N=12$ & $10^{-4}$ & $|{-}\rangle$ & $7.9\!\times\!10^{3}$ & 8.93$^{\rm t}$ & 0.83$^{\rm t}$ &  & 1.46$^{\rm t}$ & \textbf{0.83} & 6.89 \\
 & $10^{-6}$ & $|{-}\rangle$ & $5.2\!\times\!10^{4}$ & 8.76$^{\rm t}$ & 0.38$^{\rm t}$ &  & 3.64$^{\rm t}$ & \textbf{0.38} & 6.50 \\
\hline
Mixed-field Ising 2D, $3\times3$ & $10^{-4}$ & tilted & $1.4\!\times\!10^{5}$ & 14 & 1.13$^{\rm t}$ &  & 1.82 & \textbf{1.13} & 2.47 \\
 & $10^{-6}$ & $|{-}\rangle$ & $7.9\!\times\!10^{6}$ & 1.18 & 0.22$^{\rm t}$ &  & -- & \textbf{0.22} & 2.06 \\
\hline
Mixed-field Ising 2D, $3\times4$ & $10^{-4}$ & tilted & $2.5\!\times\!10^{5}$ & 12 & 2.18$^{\rm t}$ &  & 2.58 & \textbf{2.18} & 4.88 \\
 & $10^{-6}$ & $|{-}\rangle$ & -- & -- & -- &  & -- & -- & -- \\
\hline
Hubbard, $L=4$ & $10^{-4}$ & RHF & $6.6\!\times\!10^{4}$ & 2.90$^{\rm r}$ & 1.03$^{\rm r}$ &  & 3.55$^{\rm r}$ & \textbf{1.03} & 1.26 \\
 & $10^{-6}$ & RHF & $1.1\!\times\!10^{5}$ & 1.98$^{\rm r}$ & 0.75$^{\rm r}$ &  & 2.87$^{\rm r}$ & \textbf{0.75} & 1.40 \\
\hline
Hubbard, $L=6$ & $10^{-4}$ & RHF & $2.1\!\times\!10^{6}$ & 6.19$^{\rm r}$ & 1.76$^{\rm r}$ &  & -- & \textbf{1.76} & 1.36 \\
 & $10^{-6}$ & RHF & $1.0\!\times\!10^{7}$ & 2.92$^{\rm r}$ & 0.66$^{\rm r}$ &  & -- & \textbf{0.66} & 1.42 \\
\hline
H$_4$, 1.5 \AA & chem.\ acc. & RHF & $1.6\!\times\!10^{5}$ & 4.64$^{\rm c}$ & 6.78 & 3.04 & 3.69$^{\rm i}$ & \textbf{3.04} & 0.67 \\
 & $10^{-4}$ & RHF & $2.7\!\times\!10^{5}$ & 3.76$^{\rm c}$ & 7.13 & 19 & 7.37 & \textbf{3.76} & 0.80 \\
\hline
H$_4$, 2.0 \AA & chem.\ acc. & RHF & $2.3\!\times\!10^{5}$ & 3.81 & 7.89 & 2.71$^{\rm i}$ & 2.57$^{\rm i}$ & \textbf{2.57} & 0.59 \\
 & $10^{-4}$ & RHF & $6.5\!\times\!10^{5}$ & 2.88 & 5.55 & 4.06 & 3.45$^{\rm i}$ & \textbf{2.88} & 0.76 \\
\hline
LiH, 1.6 \AA & chem.\ acc. & RHF & $1.2\!\times\!10^{5}$ & 2.72 & 5.10$^{\rm c}$ & 7.81 & 5.84$^{\rm i}$ & \textbf{2.72} & 1.51 \\
 & $10^{-4}$ & RHF & $1.6\!\times\!10^{5}$ & 3.01$^{\rm c}$ & 6.25$^{\rm c}$ & 8.73 & 5.99 & \textbf{3.01} & 1.84 \\
\hline
LiH, 3.0 \AA & chem.\ acc. & RHF & $4.0\!\times\!10^{5}$ & 2.28 & 14$^{\rm c}$ & 2.25 & 2.99 & \textbf{2.25} & 6.04 \\
 & $10^{-4}$ & RHF & $8.0\!\times\!10^{5}$ & 1.97 & 6.83$^{\rm c}$ & 1.12 & 1.49 & \textbf{1.12} & 4.38 \\
\hline
H$_6$, 1.5 \AA & chem.\ acc. & RHF & $8.6\!\times\!10^{6}$ & 5.96$^{\rm c}$ & 66$^{\rm c}$ & 60 & 13 & \textbf{5.96} & 0.51 \\
 & $10^{-4}$ & RHF & $1.5\!\times\!10^{7}$ & 5.51$^{\rm c}$ & 119$^{\rm c}$ & 128 & 240 & \textbf{5.51} & 0.72 \\
\hline\hline
TFIM 1D, $N=12$ & $10^{-4}$ & $|{-}\rangle$ & $3.3\!\times\!10^{5}$ & 1.38 & 0.17 &  & 0.36 & \textbf{0.17} & 0.96 \\
 & $10^{-6}$ & $|{-}\rangle$ & $5.9\!\times\!10^{5}$ & 1.21 & 0.10 &  & 0.45 & \textbf{0.10} & 0.98 \\
\hline
TFIM 2D, $3\times3$ & $10^{-4}$ & $|{-}\rangle$ & $5.9\!\times\!10^{5}$ & 3.83 & 0.20 &  & 0.10 & \textbf{0.10} & 2.15 \\
 & $10^{-6}$ & $|{-}\rangle$ & $4.1\!\times\!10^{6}$ & 1.49 & 0.13 &  & 0.17 & \textbf{0.13} & 1.33 \\
\hline
TFIM 2D, $3\times4$ & $10^{-4}$ & $|{-}\rangle$ & $1.5\!\times\!10^{7}$ & 0.68 & 0.13 &  & 0.02 & \textbf{0.02} & 1.26 \\
 & $10^{-6}$ & $|{-}\rangle$ & -- & -- & -- &  & -- & -- & -- \\
\end{tabular}

\end{ruledtabular}
\end{table*}

\begin{figure*}[p]
\centering
\includegraphics[width=\textwidth]{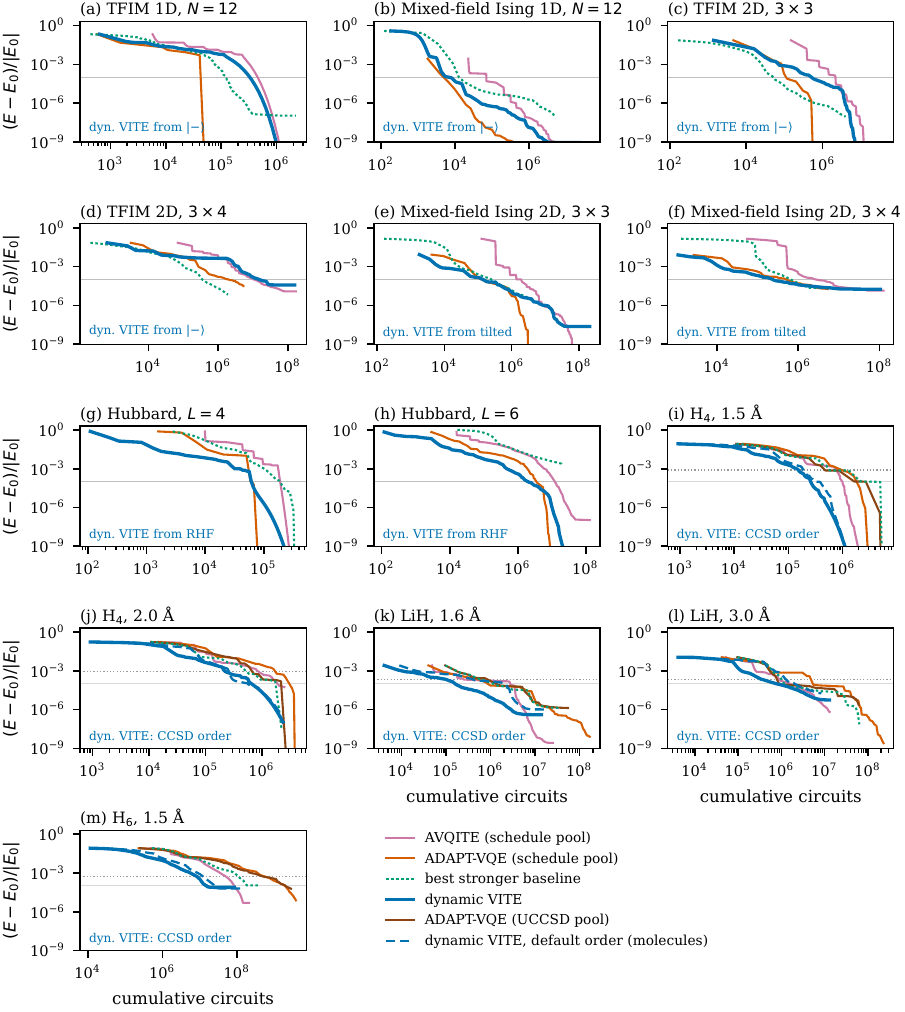}
\caption{Relative energy error versus cumulative circuits for all thirteen systems, every method from its best start. (a)--(h)~Lattice models; the gray line marks the target $10^{-4}$, and the start of dynamic VITE ($\epsilon=0.9$) is labeled. (i)--(m)~Molecules, all from RHF; the dotted line marks chemical accuracy and the gray line $10^{-4}$. Dynamic VITE is shown with the CCSD-ordered schedule (solid, Sec.~\ref{sec:molecules}) and with the default schedule (dashed). AVQITE and ADAPT-VQE use the schedule as pool (for molecules the cheaper of the default and the CCSD-ordered schedule); ADAPT-VQE with the UCCSD pool is shown for molecules; the dotted green curve is the cheapest stronger baseline at the target. Of seeded ensembles the run with the median cost is drawn.}
\label{fig:models}
\end{figure*}

Table~\ref{tab:summary} answers our question in its column ``cheapest'': the circuits needed by the cheapest adaptive method, each method from its best start, relative to those of dynamic VITE. Dynamic VITE is cheaper on seven systems: at $10^{-4}$ by 2.2 and 1.8 times on the $3\times4$ mixed-field torus and the six-site Hubbard chain, and at chemical accuracy by 2.3--6.0 times on the five molecules. It is 0.83--1.13 on the mixed-field ring, the $3\times3$ mixed-field torus and the four-site Hubbard chain, where the costs are comparable. It is 0.02--0.17 on the transverse-field models, where the adaptive methods are cheaper. Figure~\ref{fig:models} shows the convergence behind these numbers for all thirteen systems.

\emph{Which start suits which method.} The best starts differ between methods and systems. On the mixed-field models the tilted product state, which carries the magnetization of the ground state, lowers the relative error of the start from 14--39\% to 0.3--0.8\%. It helps ADAPT-VQE most, by factors of 2.3--200 at $10^{-4}$. AVQITE, whose cost is dominated by screening, is cheaper from $|{-}\rangle^{\otimes N}$ on the tori, and dynamic VITE gains little, because its first layer performs the tilt. On the transverse-field tori $|{-}\rangle^{\otimes N}$ is best for every method, because the tilted state breaks the spin-flip symmetry of the ground state and has a smaller overlap with it. On the Hubbard chains, RHF is best for dynamic VITE, AVQITE and ADAPT-VQE with the schedule pool. The stronger baselines differ: ADAPT-VQE with the GSD pool prefers the N\'eel state, and the methods with the qubit-excitation pool prefer UHF or N\'eel, because at the RHF state every gradient of that pool vanishes. Ref.~\cite{SM} gives every method from every start.

\emph{Mixed-field Ising models.} On the $3\times4$ torus the cheapest adaptive method is ADAPT-VQE with the schedule pool from the tilted state. On the $3\times3$ torus ADAPT-VQE from the tilted state needs a median of 1.13 times the circuits of dynamic VITE at $10^{-4}$, and 0.96--2.02 times over seeds. On the ring it is 17\% cheaper than dynamic VITE in every seed. The stronger baselines, which use single-string pools, need 1.5--2.6 times the circuits of dynamic VITE on the three mixed-field systems. They need many operators (26--120 to reach $10^{-4}$, against 3--10 layers for dynamic VITE), and on the tori none of them reaches $10^{-6}$ within 150 operators.

\emph{Hubbard chains.} The RHF start changes the picture most. With the hopping term as parent, the schedule is built from the commutator of hopping and interaction, and the symmetry group doubles because the spin flip is now a symmetry of $H_0$. Dynamic VITE reaches $10^{-4}$ on the four-site chain with 25 times fewer circuits than from the N\'eel state, and on the six-site chain, where it stalled at $2\times10^{-3}$ from the N\'eel state, it now reaches $10^{-6}$. On the four-site chain ADAPT-VQE from RHF matches it at $10^{-4}$ (median 1.03 times its circuits, 0.85--1.31 over seeds), and AVQITE and the best stronger baseline need 2.9 and 3.6 times as many. On the six-site chain AVQITE needs 6.2 times as many circuits as dynamic VITE, and none of the stronger baselines reaches $10^{-4}$ within 150 operators.

\subsection{Molecules}\label{sec:molecules}

For the molecules, dynamic VITE uses the CCSD-ordered schedule (Sec.~\ref{sec:schedule}). The adaptive methods receive the same classical information: AVQITE and ADAPT-VQE may use the CCSD-ordered schedule as pool, and ADAPT-type methods with the UCCSD pool may start each new operator from its CCSD amplitude instead of zero. Each method is reported with whichever choice is cheaper. Dynamic VITE reaches chemical accuracy on all five molecules, and the cheapest adaptive method needs 2.3--6.0 times as many circuits (Table~\ref{tab:summary}); at $10^{-4}$ it needs 1.1--5.5 times as many. The CCSD order lowers the cost of dynamic VITE by up to 13 times on LiH, where the default order front-loads excitations out of the Li 1s orbital, and changes it by less than 25\% on H$_4$ (Sec.~S15 of Ref.~\cite{SM}). The same information helps the adaptive methods less. At $10^{-4}$ the CCSD-ordered pool helps AVQITE on three of the five molecules and ADAPT-VQE on the two LiH geometries and on H$_6$. Starting the UCCSD operators from their CCSD amplitudes changes the cost of ADAPT-type methods by factors of 0.7--1.4, with no consistent gain. The closest competitor is ADAPT-VQE with the UCCSD pool on LiH at 3.0~\AA, which needs six operators and 1.12 times the circuits of dynamic VITE at $10^{-4}$. On H$_6$, AVQITE with the CCSD-ordered pool needs 6.0 times the circuits of dynamic VITE at chemical accuracy, the best stronger baseline (AVQITE with the QEB pool) 13 times, and the ADAPT-type methods 60--319 times. At $10^{-4}$ the ADAPT-type methods need 119--240 times, and none of the QEB baselines reaches $10^{-4}$ within 150 operators.

Every layer of the molecular schedules is a single excitation class of 2 or 8 strings from the first commutator level. On H$_4$ the ADAPT-type methods need about as many operators (10--19) as dynamic VITE needs layers (13--21), but every growth step brings a full re-optimization. On LiH they need only 5--10 UCCSD operators, whereas dynamic VITE appends 27--59 layers. There the trigger grows the circuit further than needed. Close to chemical accuracy it appends 23 and 41 layers in a single growth event, because none of them lowers $\rho$ below $\epsilon$. On LiH at 3.0~\AA\ a fixed depth of 10 layers, chosen in hindsight, reaches chemical accuracy with 0.62 times the circuits of dynamic VITE, and on H$_4$ and LiH the best fixed depth costs 0.46--0.94 times as much (Sec.~S14 of Ref.~\cite{SM}). Limiting the layers appended per growth event could therefore make dynamic VITE cheaper on the molecules.

\subsection{Origin and limits of the advantage}\label{sec:limits}

\emph{Origin.} Dynamic VITE spends nothing on choosing operators: 92--99\% of its circuits go into the metric on the spin models and the Hubbard chain, and 56\% into the force on H$_4$. AVQITE estimates the same McLachlan system but spends 83--91\% of its circuits on screening on these lattice models. ADAPT-VQE never estimates the metric, but 70--93\% of its circuits go into the BFGS re-optimizations, and batching (TETRIS) saves screening rounds, not optimizer evaluations; TETRIS spends 80--99\% on re-optimization (Sec.~S12 of Ref.~\cite{SM}). The cost of an adaptive method is therefore set by how many operators it needs. A symmetry-class layer moves many strings with one parameter in a direction the commutators single out, which keeps the number of layers small where the symmetry group is large or the parent is good.

\emph{Transverse-field Ising models.} Here ADAPT-VQE needs few operators (6 to reach $10^{-4}$ on the $3\times3$ torus) and is 5--8 times cheaper than dynamic VITE. The single-string baselines are cheaper still: TETRIS-ADAPT-VQE needs 0.10 and 0.02 times the circuits of dynamic VITE on the $3\times3$ and $3\times4$ tori, and AVQITE with the qubit pool 0.36 times on the ring. A longitudinal-field scan on the $3\times3$ torus (Sec.~S11 of Ref.~\cite{SM}) shows the crossover: with each method from its best start, the cheapest adaptive method needs 0.18 and 0.58 times the circuits of dynamic VITE at $h_z=0.1$ and 0.25, and 1.13, 1.27 and 2.2 times at $h_z=0.5$, 0.75 and 1.

\emph{Tight targets.} The cost of the metric grows quadratically with depth. At $10^{-6}$, ADAPT-VQE from its best start is 2.6--4.5 times cheaper than dynamic VITE on the mixed-field ring and $3\times3$ torus and 1.3--1.5 times cheaper on the Hubbard chains, while on the molecules dynamic VITE stays cheaper at $10^{-4}$ (Sec.~\ref{sec:molecules}). On the $3\times4$ tori no method reaches $10^{-6}$ within the schedule length or 150 operators.

\emph{Weak layers.} Where the schedule front-loads weak layers, as on the $3\times4$ transverse-field torus and the Hubbard chains from the N\'eel state, dynamic VITE walks through them while screening skips them. On LiH a run of weak layers makes it append more layers than needed (Sec.~\ref{sec:molecules}). A better parent removes much of this problem on the Hubbard chains. Hybrids that screen only when the residual signals a weak layer help there but cost too much where the schedule works. A cheaper hybrid screens only the next five scheduled layers at each growth event. On the four-site Hubbard chain from RHF it needs 0.56 times the circuits of dynamic VITE at $10^{-4}$, which also makes it cheaper than every adaptive method there; on the other seven tested systems and starts it costs 0.37--2.8 times as much as dynamic VITE at $10^{-4}$ (Sec.~S16 of Ref.~\cite{SM}).

\section{Discussion}\label{sec:discussion}

\emph{When to use dynamic VITE.} Three conditions favor it. First, the schedule should be short: the symmetry group should be large, the parent close to the ground state, or, for molecules, the first level ordered by classical amplitudes. Second, the ground state should carry correlations that adaptive methods must assemble from many operators, as in the mixed-field models with broken spin-flip symmetry, the Hubbard chains and the molecules. Where few operators suffice, as in the transverse-field models, selection is cheap and adaptive methods win. Third, the target should be moderate, around $10^{-4}$ or chemical accuracy. These conditions rest on thirteen systems of at most 16 qubits and are guidelines, not guarantees.

\emph{The role of the start.} The parent operator links the start to the schedule. A better start therefore shortens the schedule as well as the evolution. On the mixed-field models the tilted start gives every method the magnetization for free, and what remains for dynamic VITE is the cost of the correlations beyond the product state. On the Hubbard chains the RHF parent turns the first commutator level into the commutator of hopping and interaction, which the N\'eel parent does not provide. Choosing the start per method avoids comparing dynamic VITE against handicapped baselines. It also shows that the best start is not always the one with the lowest energy: on the Hubbard chains the schedule-pool methods prefer RHF to the lower-energy UHF state, and the qubit-excitation pool cannot leave the RHF state at all.

\emph{Cost of the metric.} The metric dominates the cost of dynamic VITE. Every scheduled layer is symmetric, so symmetry-equivalent entries of $M$ and $V$ are equal; measuring them once lowers the cost on the mixed-field ring by factors of 8--18 at $10^{-4}$ (Sec.~S10 of Ref.~\cite{SM}), a saving single-string pools cannot use. All costs above are without this reduction. Approximating $M$ further would break the bound $0\le\rho\le1$ on which the trigger relies.

\emph{Limitations.} All systems have at most 16 qubits and are simulated exactly, without noise. The cost model counts circuits, not shots or gates. ADAPT-VQE uses exact gradients and BFGS, which favors it. The stronger baselines stop at 150 operators; with more operators they might reach $10^{-4}$ on the six-site Hubbard chain and, with the QEB pool, on H$_6$. Overlap-guided and coupled-exchange variants of ADAPT-VQE~\cite{Feniou2023,Ramoa2025} were not tested. The CCSD order was tested on five molecular systems; where CCSD itself fails, its amplitudes may also give a poor order. We have not identified a quantity that predicts how many layers the schedule needs.

\section{Conclusion}\label{sec:conclusion}

A fixed operator order can reach the accuracy of adaptive screening at a comparable or lower measurement cost. In dynamic VITE the order is built from nested commutators of the Hamiltonian with a parent operator of the start, and the circuit grows when the McLachlan residual signals that it has stalled. With every method started from the classically computed state that suits it best, dynamic VITE needed 1.8--6.0 times fewer circuits than the cheapest adaptive variant we tested on seven of thirteen systems at $10^{-4}$ or chemical accuracy, and it was within 20\% of the cheapest on three more. Classical calculations before the quantum run, a mean-field start with its parent and, for molecules, a coupled-cluster order, take over part of the selection that adaptive methods perform on the device. Dynamic VITE is more expensive where adaptive methods need few operators and at tight targets.

\begin{acknowledgments}
This work was supported by the National Science and Technology Council (Taiwan) under grant NSTC 114-2112-M-182-002-MY3 and by Chang Gung Memorial Hospital under grant BMRPL94.
\end{acknowledgments}

\section*{Code availability}
The Python code is available at \url{https://github.com/plain12356/Dynamic-VITE}. 

\bibliography{refs}
\end{document}